# Piggy-backing protein domains with Formal Concept Analysis

Susan Khor
September 3, 2013

**Abstract**

Identifying reliable domain-domain interactions (DDIs) will increase our ability to predict novel protein-protein interactions (PPIs), to unravel interactions in protein complexes, and thus gain more information about the function and behavior of genes. One of the challenges of identifying reliable DDIs is domain promiscuity. Promiscuous domains are domains that can occur in many domain architectures and are therefore found in many proteins. This becomes a problem for a method where the score of a domain-pair is the ratio between observed and expected frequencies because the PPI network is sparse. As such, many protein-pairs will be non-interacting and domain-pairs with promiscuous domains will be penalized. This domain promiscuity challenge to the problem of inferring reliable DDIs from PPIs has been recognized, and a number of work-arounds have been proposed. In this paper, we report an application of Formal Concept Analysis (FCA) to this problem. We find that the relationship between formal concepts provide a natural way for rare domains to elevate the rank of promiscuous domains, and enrich highly ranked domain-pairs with reliable DDIs. This piggy-backing of promiscuous domains onto rare domains is possible due to the domain architecture of proteins which mixes promiscuous with rare domains.

## 1. Introduction

Proteins comprise domains which are evolutionary conserved sequence segments with the ability to fold and function independently. An important class of domains mediates protein-protein interactions (PPIs); although not all interactions between proteins can be attributed to interactions between domains, and not all domains in multi-domain proteins play a direct role in protein interaction. Nonetheless, many computational methods which seek to predict PPIs with high accuracy rely on computationally inferred domain-domain interactions (DDIs), e.g. [1].

Ideally, the inferred DDIs used to support the predicted PPIs are highly reliable themselves, that is there is a large overlap between the set of inferred DDIs and the set of physically verified or gold standard domain-domain interactions (GDDIs). This ideal is desirable not only to tease out specific interactions in a protein complex, but also to give predictive power to protein-protein interaction prediction methods (more on this point later in this section).

However, using GDDIs to predict PPIs generates a large number of false positives (non-interacting protein-pairs predicted as interacting) and thus reduces the accuracy of the prediction method. The large number of false positives stem from the fact that GDDIs are enriched with *promiscuous domains*. Promiscuous domains can occur in many domain architectures [2] and thus appear in many proteins. But since the PPI network is sparse, many of these protein-pairs will be non-interacting.

It is parsimonious to re-use domain-pairs that can interact to facilitate PPIs. Indeed, many DDIs are conserved by evolution [3] and there is a high degree of DDI re-use by PPIs [4]. Promiscuous domains are observed to be heavily involved in PPI mediation [2]. In theory, PPI prediction methods which depend on inferred DDIs rely on the presence of this parsimony. The basic underlying thinking is DDIs inferred from PPIs in the training set can then be used to predict PPIs in the test set. Fundamental to the success of this strategy is a commonality between the proteins in the training and test sets, at least in the form of domain-pairs. When this commonality is reduced, e.g. through the use of rare DDIs to predict PPIs, the power (ability to generalize from sample to population) of a prediction method weakens. This flaw in existing computational PPI prediction methods was demonstrated in [5] wherein the predictive performances of seven PPI prediction methods deteriorated significantly as the intersection between the training protein set and the test protein set decreased to null (the number of domain-pairs in common also decreased Appendix:SM1).

The 'drift towards rare domain-pairs' phenomenon in PPI prediction methods has been noted [6]. Such rare domain-pairs comprise domains which occur infrequently in a given protein sample but occurs in interacting protein-pairs so that rare domain-pairs appear to be highly reliable DDIs and good indicators of putative PPIs (since they dampen the increase in false positives). However, rare domain-pairs are often *not* GDDIs. Further, rare domain-pairs have weak predictive value since by their nature, they are not commonly found in proteins and therefore the information that they interact is less reusable for the purpose of predicting PPIs. We suggest that the 'drift towards rare domain-pairs' phenomenon is partly a consequence of how computational PPI prediction methods are evaluated. However, the 'drift towards rare domain-pairs' is also because *promiscuity prevents GDDIs from being highly ranked* in computational methods to infer DDIs. Recognizing this domain promiscuity problem, additional measures have been taken to counteract its effects when inferring DDIs from PPIs, e.g. [6-8].

In this paper, we show how Formal Concept Analysis (FCA) [8] (section 4) can be used to overcome the promiscuity problem for detecting GDDIs from a given set of PPIs. Our method is different from previous proposals in several ways:

(i) It is a more discrete approach, and we believe this is the first use of FCA in this manner.
(ii) We are able to identify necessary conditions for our method to work.
(iii) We can relate the results of our method to the characteristics of the given input data.

**2. Basic definitions**

Let $\mathcal{P}$ be the set of proteins and $\mathcal{D}$ the set of domains. Every protein in $\mathcal{P}$ comprises one or more domains in $\mathcal{D}$. $D(x)$ denotes the finite set of domains for protein $x$. If $|D(x)| = 1$, $x$ is a single-domain protein; if $|D(x)| > 1$, $x$ is a multi-domain protein. Every domain in $\mathcal{D}$ is contained within one or more proteins in $\mathcal{P}$. The set of proteins which contains domain $a$ is $P(a) = \{ x \mid (\forall x \in \mathcal{P})\ a \in D(x) \}$. The frequency of domain $a$ in $\mathcal{P}$ is $N(a) = |P(a)|$. $x = D(x) = \{a, b, c\}$ where $x \in \mathcal{P}$ and $\{a, b, c\} \subset \mathcal{D}$ denotes protein $x$ is its set of domains $D(x)$ which in turn comprises domains $a$, $b$ and $c$.

The set of PPIs is a relation on $\mathcal{P}$. This relation is symmetric, i.e. if $(x, y)$ is an interacting protein-pair, then so is $(y, x)$. It is possible for proteins to self-interact. Let $(x, y)^1$ denote $(x, y) \in$ the set of PPIs. $(x, y)^1$ implies proteins $x$ and $y$ come from the same organism, i.e.: $\mathbb{O}(x) = \mathbb{O}(y)$. The set of non-PPIs is also a symmetric relation on $\mathcal{P}$, and a non-interacting protein-pair $(x, y)^0$ also implies $\mathbb{O}(x) = \mathbb{O}(y)$. $\mathbb{O}(x) = \mathbb{O}(y)$ implies either $(x, y)^1$ or $(x, y)^0$. There may be pairs in $\mathcal{P} \times \mathcal{P}$ which are neither interacting nor non-interacting because they do not satisfy the same organism condition.

The set of DDIs is a symmetric relation on $\mathcal{D}$, and domain self-interaction is possible. A protein-pair $(x, y)$ generates domain-pairs, each of which may or may not be reliable, through the cross-product of their domains, i.e.: $D(x) \times D(y)$. A domain-pair $(a, b)$ generates a set of protein-pairs, each of which may or may not be interacting, through the cross-product of their respective protein sets, i.e.: $P(a) \times P(b)$. For a domain-pair $(a, b)$ to be a DDI, it must generate at least one interacting protein-pair.

## 3. The Riley dataset and its characteristics

The Riley dataset [8] has been re-used in a number of studies, e.g.: [6, 10]. This dataset comprises 11,403 proteins from 68 organisms (Table 1). The proteins are associated with 12,455 Pfam-A and Pfam-B domains. Amongst the set of proteins are 26,032 protein-protein interactions (PPIs). Interactions and non-interactions between protein-pairs are restricted to proteins from the same organism [8]. The interaction of two proteins $x$ and $y$ implies interactions between the domains of $x$ and the domains of $y$. The Riley set of proteins, domains and PPIs generate 177,233 putative domain-domain interactions (DDIs).

Amongst these possible DDIs are 783 gold standard domain-domain interactions (GDDIs). GDDIs are domain-pairs whose physical interaction has been verified experimentally. The GDDIs were obtained from [6]. Over half (403/783 = 51.57%) of the GDDIs are self-interacting (homotypic), but less than 1% (1262/176450) of the non-GDDIs are self-interacting. DDIs which mediate PPIs are enriched with homotypic domain-pairs [3, 11]. A PPI with at least one GDDI is a gold-PPI (GPPI). There are 850 GPPIs in the Riley dataset.

**Table 1** A sample of organisms in the Riley dataset.

| Organism | Number of proteins in the Riley dataset |
|---|---|
| *Drosophila melanogaster* | 3777 |
| *Saccharomyces cerevisiae* | 3476 |
| *Caenorhabditis elegans* | 2233 |
| *Homo sapien* | 799 |
| *Schizosaccharomyces pombe* | 10 |
| *Bacillus subtilis* | 9 |

In the remainder of this section, we examine the Riley dataset to support assertions made in Section 1 and the discussion in the rest of this paper. Specifically, the data characteristics of interest are:
(i)     Highly reliable domain-domain interactions (GDDIs) are enriched with promiscuous domains.

(ii) GDDIs generate significantly more true positive PPIs and also more false positive PPIs than non-GDDIS. More true positive PPIs agrees with the parsimony or the re-use principle for GDDIs, and more false positive PPIs accords with the promiscuity of gold domains.

(iii) Protein domain architectures are mostly a mix of rare and promiscuous domains.

**3.1 GDDIs are significantly enriched with promiscuous domains.**

To test this assertion and its corollary that domain-pairs comprising rare domains are often not GDDIs, we pooled the domain partners of all GDDIs and compared their frequency of occurrence against the frequency of occurrence of all 12,455 domains. If *the set of gold domains is significantly more promiscuous than the set of all domains,* the assertion is confirmed and we conclude that GDDIs are significantly enriched with promiscuous domains. A *gold domain* is a domain that participates in at least one gold standard domain-domain interaction (GDDI). There are 642 gold domains in the Riley dataset. A domain is more promiscuous if it occurs more frequently in a given set of proteins, i.e. given $\mathcal{P}$ and $\{a, b\} \subset \mathcal{D}$, $N(a) > N(b)$ implies *a* is more promiscuous than *b*.

We observed that domains are not distributed normally amongst proteins. Few domains occur much more frequently and most domains occur infrequently. The log-log plot in Fig. 1 shows the right-skewed distribution of domain occurrence which is exhibited more clearly by the set of all domains than the set of gold domains (even though the set of all domains is much larger than the set of gold domains). The difference in frequency distributions is significant. Analysis with R's Wilcox test function confirms that the set of gold domains is significantly more promiscuous than the set of all domains (Appendix SM2).

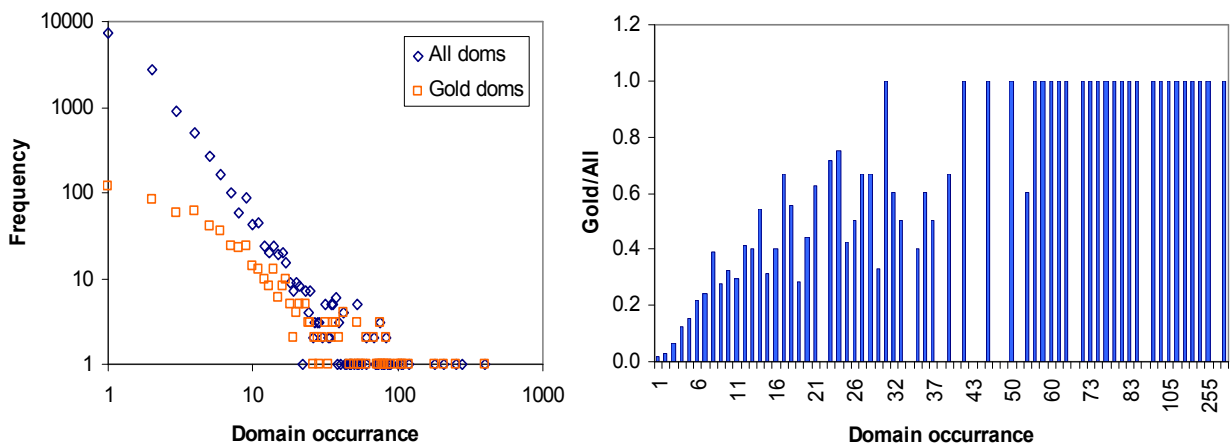

**Fig. 1** (Left) Domains do not occur with the same frequency; there are rare and promiscuous domains. The rare domains outnumber the promiscuous domains several fold. There are fewer rare domains in the set of gold domains than in the set of all domains. (Right) In general, the more frequently a domain occurs, the more likely it is to be a gold domain.

The bar chart in Fig. 1 gives the ratio of gold domains against all domains for every domain occurrence value. Gold/All = 0 for a domain occurrence value of $x$ means no gold domain occurred $x$ times (the domains which occurred that many times are all not gold domains). Gold/All = 1 for a domain occurrence value of $x$ means all the domains which occurred $x$ times are gold domains. Observe the tendency of Gold/All to equal 1 at larger domain occurrence values. This supports the assertion that GDDIs are significantly enriched with promiscuous domains.

**3.2 GDDIs generate significantly more true positive and more false positive PPIs than non-GDDIs.**
To test this assertion, the set of all interacting protein-pairs (PPIs) and the set of all non-interacting protein-pairs (non-PPIs) were generated for each of the 177,233 putative domain-domain interactions (DDIs). A DDI is either a GDDI or a non-GDDI. The set of DDIs comprise 783 GDDIs and 176,450 non-GDDIs. If the GDDIs generate significantly larger sets of PPIs than the non-GDDIs, we conclude that GDDIs generate significantly more true positive PPIs than non-GDDIs. If the GDDIs generate significantly larger sets of non-PPIs than the non-GDDIs, we conclude that GDDIs generate significantly more false positive PPIs than non-GDDIs.

A domain-pair ($a$, $b$) generates a set of protein-pairs, each of which may or may not be interacting, through the cross-product of their protein sets, i.e. P($a$) × P($b$). A PPI prediction for a protein-pair ($x$, $y$) predicts that protein $x$ interacts with protein $y$. This prediction is a true positive if $(x, y)$[1] can be found in the given set of PPIs and a false positive otherwise.

We use R's `t.test` (non-homogeneous variance) and `Wilcox.test` to compare the sizes of the PPI sets generated by GDDIs and by non-GDDIs, and the sizes of the non-PPIs sets produced by GDDIs and by non-GDDIs. The statistical tests confirm that GDDIs generate significantly larger sets of PPIs than non-GDDIs (Fig. SM3.1) and GDDIs generate significantly larger sets of non-PPIs than non-GDDIs (Fig. SM3.2). These results allow us to conclude that GDDIs generate significantly more true positive and more false positive PPIs than non-GDDIs. More true positive PPIs agrees with the parsimony or the re-use principle for GDDIs, and more false positive PPIs accords with the promiscuity of gold domains.

**3.3 Protein domain architectures are mostly a mixture of rare and promiscuous domains.**
Promiscuous domains are those which occur frequently. Rare domains are those which occur infrequently. A domain is classified as either promiscuous or rare. While a rare domain may occur in only a handful of proteins, collectively the set of rare domains (at threshold |P($a$)| ≤ 5) occur in over 70% of the proteins in the Riley dataset (Table 2). Table 2 shows the breakdown of proteins by domain architecture. The most popular domain architecture is a combination of promiscuous and rare domains. At least a third of the proteins have the mixed domain architecture. Mixed architecture proteins are the workhorse proteins in our method. They enable the piggy-back mechanism described in section 5. When $|P(a)| \leq 5$ is the criterion for identifying rare domains, the domains pooled from the set of mixed architecture proteins include 53.43% of the gold domains (Table 3). When $|P(a)| \leq 10$, the number of proteins with mixed domain architecture decreases, resulting in a concomitant decline in the coverage of gold domains by the set of domains from mixed architecture proteins to 41%.

**Table 2** Breakdown of proteins by domain architecture. The most popular domain architecture is a combination of promiscuous and rare domains. At least a third of the proteins have this mixed architecture. $|P(a)| \leq 5$ is a threshold for identifying rare domains. At this threshold, domain *a* is a rare domain if there are at most 5 proteins containing domain *a*. Otherwise, domain *a* is a promiscuous domain. The Riley dataset has 12,455 domains and 11,403 proteins, of which 4,541 are single-domain proteins.

| Domain architecture | Protein type | Number and % of proteins | |
|---|---|---|---|
| | | $|P(a)| \leq 5$<br>11,696 rare domains<br>759 promiscuous | $|P(a)| \leq 10$<br>12,148 rare domains<br>307 promiscuous |
| Only promiscuous domains | Single-domain | 2,183 (19.14%) | 1,706 (14.96%) |
| | Multiple-domain | 905 (7.94%) | 420 (3.68%) |
| Only rare domains | Single-domain | 2,358 (20.68%) | 2,835 (24.86%) |
| | Multiple-domain | 1,737 (15.23%) | 2,534 (22.22%) |
| Mixed: promiscuous and rare domains combined | Multiple-domain | 4,220 (37.01 %) | 3,908 (34.27%) |

**Table 3** Coverage of the 642 gold domains by proteins with mixed domain architecture. At the $|P(a)| \leq 5$ threshold, proteins with mixed architecture have a combined total of 7,827 unique domains, which includes 53.43% of the 642 gold domains in the Riley dataset.

| Rare domain threshold | Domains pooled from mixed architecture proteins | Gold domains covered |
|---|---|---|
| $|P(a)| \leq 5$ | 7,827 | 343/642 = 53.43% |
| $|P(a)| \leq 10$ | 6,755 | 263/642 = 40.97% |

## 4. Formal Concept Analysis

Formal Concept Analysis (FCA) [9] is a technique to organize a (finite) set of objects *G* (German: Gegenstände) by their common attributes and dually a (finite) set of attributes *M* (German: Merkmale) by their common objects into a (finite) set of partially ordered pairs of sets called (formal) *concepts*. Implicit is a binary relation $I \subseteq G \times M$ which associates objects with attributes. $(g, m) \in I$ or equivalently $g\,I\,m$ denotes object *g* has attribute *m*. The triplet (*G*, *M*, *I*) forms the (formal) *context* within which a FCA is carried out. For small finite examples, a context can be specified completely with a cross-table. The resulting set of concepts, denoted $\mathfrak{B}(G, M, I)$ (German: $\mathfrak{B}$ for Begriff), forms a *concept lattice*.

For the application in this paper, the set of objects is the set of proteins, i.e. $G = \mathcal{P}$, the set of attributes is the set of domains, i.e. $M = \mathcal{D}$, and $g\,I\,m$ denotes protein *g* has domain *m*, i.e. $m \in D(g)$ and dually $g \in P(m)$. Table 4 is the cross-table for the relation between proteins and domains associated with the organism *S. pombe* in the Riley dataset. The concept lattice expressing this relation is given in Fig. 2.

**Table 4** A cross-table representing the relation between proteins and domains associated with *S. pombe* in the Riley dataset. E.g.: the domain set for protein 353, D(353) = {APSES, Ank, Pfam-B_39251, Pfam-B_45975}; and the protein set for domain Pkinase, P(Pkinase) = {10, 1076, 136, 16949}.

|  |  | Objects = Proteins (uid) | | | | | | | | | Domain Freq. |
|---|---|---|---|---|---|---|---|---|---|---|---|
|  |  | 10 | 1076 | 136 | 16 | 275 | 353 | 620 | 659 | 683 | 949 |  |
| Attributes = Domains | APSES |  |  |  |  |  | × |  |  |  |  | 1 |
|  | Ank |  |  |  |  |  | × |  |  |  |  | 1 |
|  | Cyclin_C |  |  |  |  |  |  | × |  |  |  | 1 |
|  | Cyclin_N |  |  |  |  |  |  | × | × |  |  | 2 |
|  | PBD |  |  |  |  |  |  |  |  |  | × | 1 |
|  | Pfam-B_106217 |  |  |  | × |  |  |  |  |  |  | 1 |
|  | Pfam-B_2441 |  |  |  |  |  |  |  |  |  | × | 1 |
|  | Pfam-B_33993 |  |  |  | × |  |  |  |  |  |  | 1 |
|  | Pfam-B_39251 |  |  |  |  |  | × |  |  |  |  | 1 |
|  | Pfam-B_45975 |  |  |  |  |  | × |  |  |  |  | 1 |
|  | Pfam-B_78326 |  |  |  | × |  |  |  |  |  |  | 1 |
|  | Pkinase | × | × | × | × |  |  |  |  |  | × | 5 |
|  | RA |  |  |  |  |  |  |  |  | × |  | 1 |
|  | Ras |  |  |  |  | × |  |  |  |  |  | 1 |
|  | SAM_2 |  |  | × |  |  |  |  |  | × |  | 2 |
| Domains per protein |  | 1 | 1 | 2 | 4 | 1 | 4 | 2 | 1 | 2 | 3 |  |

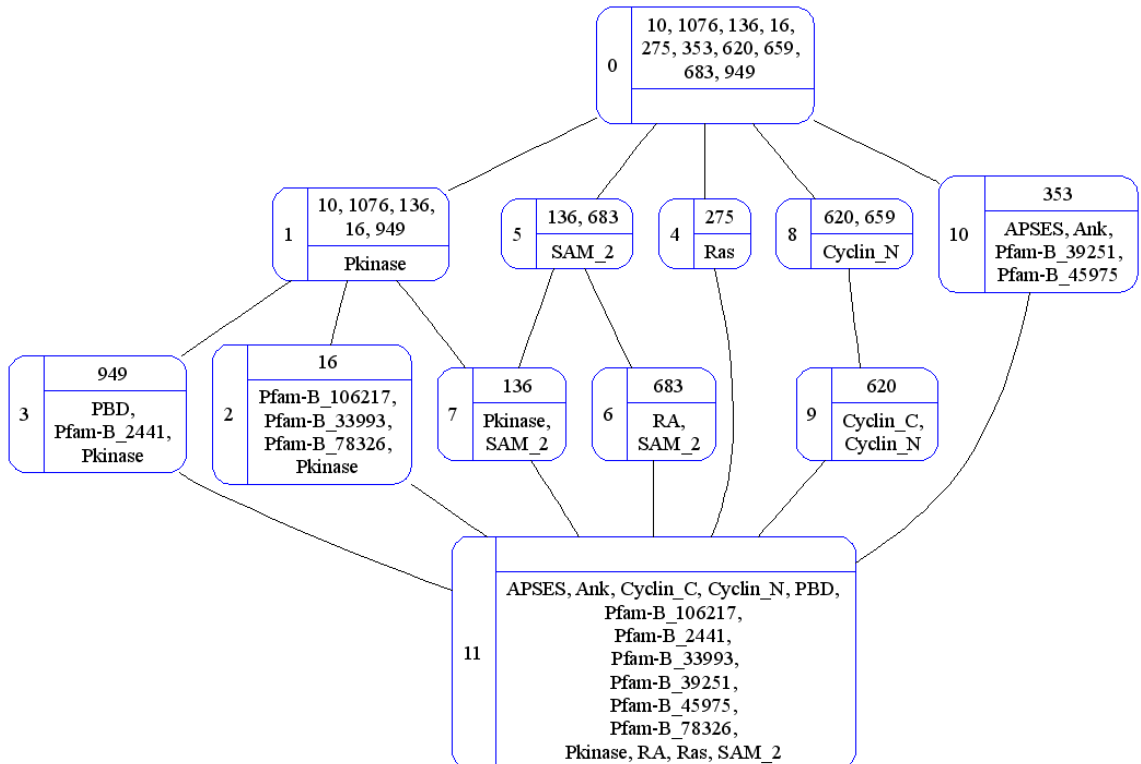

**Fig. 2** The concept lattice for the *S. pombe* relation in Table 4.

### 4.1 Details

In this section, we describe FCA in enough detail to support the discussion in this paper. The more mathematically inclined are referred to [9] for a rigorous and complete exposition of FCA.

A concept $c \in \mathfrak{B}(G, M, I)$ is an ordered pair of sets $(O, A)$ such that $O \subseteq G$, $A \subseteq M$ and the set of all attributes common to all objects in $O$ under relation $I$ is $A$ and the set of all objects with attributes in $A$ under relation $I$ is $O$. More formally, the last two conditions are $A = O' = \{m \in M \mid (\forall g \in O) \, g \, I \, m\}$ and $O = A' = \{g \in G \mid (\forall m \in A) \, g \, I \, m\}$ respectively. If this seems a bit chicken-and-egg, these last two conditions can be satisfied by working from the power-set of objects or alternatively from the power-set of attributes; but there exist more efficient FCA algorithms. We use Lindig's C implementation called Colibri-concepts [12] which is freely available on-line and runs on Linux. For the complete protein-domain relation in the Riley dataset (section 3), Colibri-concepts produced a lattice with 8,894 concepts in under 5 minutes on a Linux machine allocated with a maximum of 2Gbs of memory.

The set of objects $O$ is called the *extent* of concept $c$, and the set of attributes $A$ is known as the *intent* of concept $c$. We denote a concept's extent as $O(c)$ and its intent as $A(c)$. The prime symbol $'$ denotes the mapping from an extent to its intent and vice versa, i.e.: let a concept $c = (O, A)$, then $O' = A$, $A' = O$, $O = O''$, $A = O'''$, and so on. This pair of maps between the set of extents and the set of intents forms a Galois connection between the two partially ordered sets. The statements $O = O''$ and $A = O'''$ are true due to the maximal condition for extents and intents. This implies that if a set of objects (attributes) forms the extent (intent) of a concept, then the set of objects (attributes) uniquely identifies the concept, and conversely a concept unambiguously identifies its extent and its intent.

The set of concepts is ordered by set inclusion $\langle \mathfrak{B}(G, M, I); \leq \rangle$. For two distinct concepts in $\mathfrak{B}(G, M, I)$, $(O_1, A_1) \leq (O_2, A_2)$ implies $O_1 \subset O_2$ and dually $A_1 \supset A_2$. The join (least upper bound) and meet (greatest lower bound) are defined for every pair of non-comparable concepts in $\langle \mathfrak{B}(G, M, I); \leq \rangle$. $\langle \mathfrak{B}(G, M, I); \leq \rangle$ forms a concept lattice. Intuitively, a concept lattice is a two-in-one lattice with a right-side up lattice for the set of extents and an upside down lattice for the set of intents. More formally, a concept lattice is a complete lattice with a top element $\top = (G, \emptyset)$ and a bottom element $\bot = (\emptyset, M)$. A complete lattice defined on a subset of a power-set is closed under arbitrary joins (in the form of unions) and meets (in the form of intersections) [13]. Within a concept lattice, the join (supremum) of two arbitrary concepts $c_1 \vee c_2 = ( \, (O(c_1) \cup O(c_2))'', A(c_1) \cap A(c_2) \, )$, and the meet (infimum) of two arbitrary concepts $c_1 \wedge c_2 = ( \, (O(c_1) \cap O(c_2), (A(c_1) \cup A(c_2))'' \, )$. $c_1 \vee c_2$ is a concept since $A(c_1) \cap A(c_2) = (O(c_1) \cup O(c_2))'$ and $(O'', O')$ is always a concept. Similarly, $c_1 \wedge c_2$ is a concept since $(O(c_1) \cap O(c_2)) = (A(c_1) \cup A(c_2))'$ and $(A', A'')$ is always a concept. The intersection of any number of extents (intents) always results in an extent (intent). The same is not generally true for unions of extents (intents) [9, p.19]. Rather, $(O(c_1) \cup O(c_2)) \subseteq (O(c_1) \cup O(c_2))''$ and $(A(c_1) \cup A(c_2)) \subseteq (A(c_1) \cup A(c_2))''$ hold.

The down-set of $c$ represented as $\downarrow\{c\} = \{ x \in \langle \mathfrak{B}(G, M, I); \leq \rangle \mid x \leq c \}$. The up-set of $c$ represented as $\uparrow\{c\} = \{ x \in \langle \mathfrak{B}(G, M, I); \leq \rangle \mid c \leq x \}$. The extent of a concept $c$ is the union of the extent of each concept $\in \downarrow\{c\}$. The intent of a concept $c$ is the union of the intent of each concept $\in \uparrow\{c\}$. This relationship between concepts makes it possible to reduce the labeling of concepts to objects and attributes specific to a concept (Fig. 3). Changing the labels does not change the concepts. A concept lattice with reduced labeling is a *reduced concept lattice*.

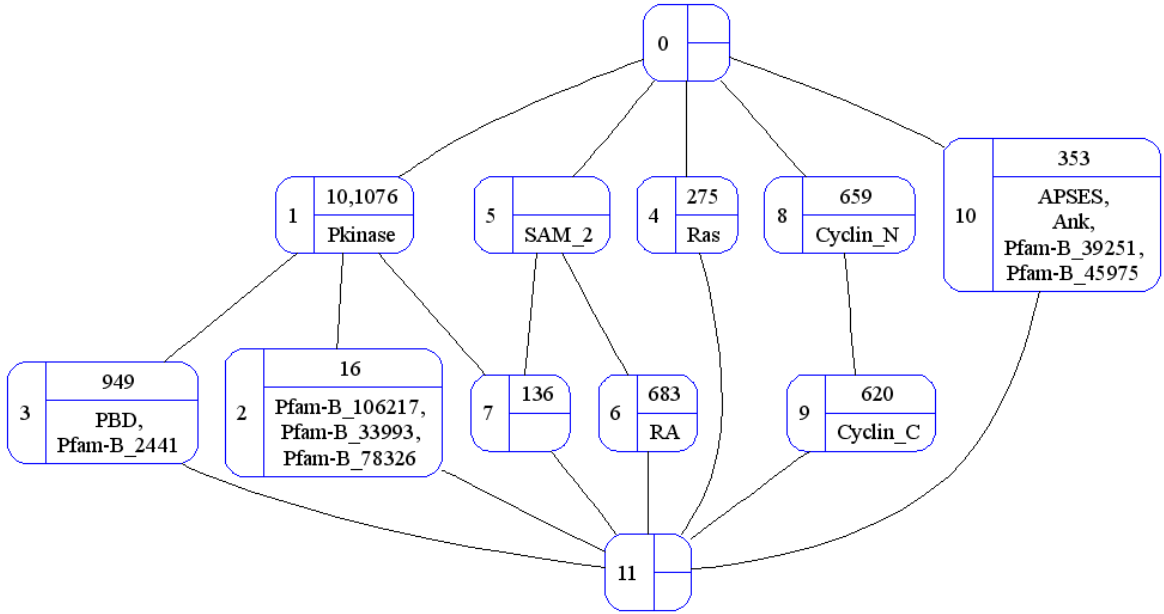

**Fig. 3** The **oa** concept lattice for the relation in Table 4. The **oA** and **Oa** concept lattices can be found in the Appendix SM5.

There are then four possible ways to label a concept lattice: (i) with complete object labels and complete attribute labels (**OA**); (ii) with reduced object labels and reduced attribute labels (**oa**); (iii) with complete object labels and reduced attribute labels (**Oa**); and (iv) with reduced object labels and complete attribute labels (**oA**). The **oa** and **oA** combinations produce *object-reduced concept lattices*. The **oa** and **Oa** combinations produce *attribute-reduced concept lattices*. We explore each combination in our work on proteins and their domains in section 5. It is possible for two different combinations to produce two different outcomes because we use the labels of objects and attributes instead of the objects and attributes themselves. We use $O^L(c)$ to refer to the set of object-labels for $O(c)$, and similarly $A^L(c)$ to refer to the set of attribute-labels for $A(c)$.

**4.2 Points of note**

(i) Promiscuous domains gravitate towards the top of an attribute-reduced concept lattice (Fig. 3 & Fig. SM5.2). This is expected since for a domain to be promiscuous, it must occur in many proteins. In FCA language, involving more proteins (objects) means a larger extent, and as one goes up in a concept lattice extents increase in size, culminating in the top element whose extent is the entire object set. For *S. pombe*, the domains with frequency $N(d) > 1$ are Pkinase (5), SAM_2 (2) and Cyclin_N(2) (Table 4) and they reside in concepts one step away from the top element and two steps away from the bottom element in Fig. 3. The rare (non-promiscuous) domains gravitate towards the bottom of an attribute-reduced concept lattice since by their rarity, rare domains command smaller extents.

(ii) The position of a protein in an object-reduced concept lattice depends on the promiscuity of its domain(s). Three of the four single-domain proteins for *S. pombe* have domains whose $N(d) > 1$, and these single-domain proteins (10, 1076 and 659) reside in concepts one step away from the top element in Fig. 3. A multi-domain protein with a combination of promiscuous and rare domains will appear in an object-reduced concept lattice with its rare domains. E.g.: protein 620 appears with domain Cyclin_C and not with Cyclin_N in Fig. 3. Also in Fig. 3, protein 16 appears with its Pfam-B domains (which tend to be rare) in concept 2 and not with the more promiscuous Pkinase in concept 1. In general therefore, proteins with rare domains will gravitate towards the bottom of an object-reduced concept lattice, and proteins with only promiscuous domains will gravitate towards the top of an object-reduced concept lattice. Further, because attribute-labels accumulate downwards in a concept lattice, concepts containing proteins with a combination of only rare domains will be found in an object-reduced concept lattice below concepts containing proteins with only one rare domain. Similarly, concepts containing proteins with a combination of only promiscuous domains will be found in an object-reduced concept lattice below concepts containing proteins with only one promiscuous domain. The reduced concept lattice in the Appendix SM4 illustrates these points more clearly.

(iii) Object-label sets and attribute-label sets may be empty in a reduced concept lattice. This follows from points (i) and (ii). The more numerous rare domains will "consume" proteins and leave fewer proteins available to promiscuous domains. So an empty object-label set $O^L(c) = \varnothing$ is more likely towards the top of an attribute-reduced concept lattice. In an object-reduced concept lattice, object labels appear exactly once and the set of object labels (proteins) is finite. Dually in an attribute-reduced concept lattice, attribute labels appear exactly once and the set of attribute labels (domains) is finite.

## 5. Domain-pair Scoring and Ranking

One of the earliest methods for detecting over-represented 'correlated sequence-signatures' e.g. domain-pairs, in a database of protein-protein interactions uses the log-odds of the ratio between observed and expected frequencies to score pairs of sequence-signatures [14]. Larger scores indicate a frequency of occurrence in the database which is higher than expected by random chance. This method was called the Association method in [1] and some subsequent papers adopted this moniker, e.g.: [6].

Specifically, the score of a domain-pair ($a$, $b$) with the Association method is AM($a$, $b$) = $\log_2 [M(a, b) / (N(a) \times N(b))]$. $M(a, b) = | \{(x, y)^1 | a \in D(x)$ and $b \in D(y)\} |$ is the number of interacting protein-pairs in the database such that $a$ is a domain of protein $x$ and $b$ is a domain of protein $y$. $N(i)$ is the number of proteins in the database that has domain $i$ in its domain architecture (section 2). The scores are negative in value with a maximum of $\log_2(1) = 0$ which is the score for domain-pairs that occur only between interacting protein-pairs, and an undefined minimum of $\log_2(0)$ which is the score for domain-pairs that occur only between non-interacting protein-pairs. Domain-pairs with larger scores are ranked more favourably (given higher rank).

The Riley dataset (section 3) comprises proteins from multiple organisms and both interacting and non-interacting proteins are restricted to those from the same organism. To handle this situation, AM($a$, $b$) = $\log_2 [M(a, b) / (M(a, b) + Z(a, b))]$. $M(a, b)$ is defined previously. $Z(a, b) = | \{(x, y)^0 | a \in D(x)$ and $b \in D(y)\} |$ is the number of non-interacting protein-pairs in the database such that $a$ is a domain of protein $x$ and $b$ is a domain of protein $y$.

### 5.1 Concept-based scoring and ranking <CB, PG>

Our *concept-based scoring* scheme also uses the log-odds ratio AM($a$, $b$) described before, but the scoring is done using pairs of concepts, excluding the top and the bottom concepts (and any other concepts with either an empty extent or an empty intent). Protein interactions and non-interactions are confined to the object-labels of a concept-pair, and a domain-pair may be evaluated zero or more times depending on which type of concept lattice is used. If more than one score exists for a domain-pair, the domain-pair is assigned the largest score and the number of unique scores strictly smaller than the largest score is recorded. If a domain-pair is scoreless (e.g. $\log_2(0)$), it is ranked below all other domain-pairs with scores. Because there is potential for many domain-pairs to have the same score, a two-tier system which considers the best score (CB) and how the best score was obtained (PG) is used to rank domain-pairs. Domain-pairs with larger CB and larger PG are ranked more favourably in terms of interaction reliability. The expectations are (i) highly ranked domain-pairs are enriched with GDDIs, and (ii) in a set of putative DDIs for a PPI, *the* highest ranked DDI is a GDDI [15].

Define $\mathcal{C}$ as the set of concepts for a context less the top and the bottom elements, i.e.: $\mathcal{C} = \mathfrak{B}(\mathcal{P}, \mathfrak{D}, I \subseteq \mathcal{P} \times \mathfrak{D}) \setminus \{\top, \bot\}$. Basic to the concept-based scoring scheme is the notion of an interaction between two concepts denoted $c_1 \times c_2$ with $c_1, c_2 \in \mathcal{C}$. The operation is commutative: $c_1 \times c_2 = c_2 \times c_1$. Also, $c_1$ may be the same concept as $c_2$. $c_1 \times c_2 = (O^L(c_1) \times O^L(c_2), A^L(c_1) \times A^L(c_2))$. The cross-product of the two sets of object-labels $O^L(c_1) \times O^L(c_2)$ produces protein-pairs which are used to generate a log-odds score for every domain-pair produced by the cross-product of the two attribute-label sets $A^L(c_1) \times A^L(c_2)$. If the concept lattice is not attribute-reduced, a domain-pair $(a, b)$ may have more than one score, and the concept-based score $CB(a, b)$ is the largest score. The number of unique scores strictly smaller than the largest score makes the PG score for $(a, b)$. The PG score for a domain-pair represents the number of times the domain-pair has improved its CB score through piggy-backing. A domain-pair without a CB score has PG = 0, and is ranked below all other domain-pairs with scores (this only happens with **oa** concept lattices). The set of concept pairs $\{(c_i, c_j)\}$ used to evaluate a domain-pair $(a, b)$ is determined by the attribute-label sets as follows: $\{(c_i, c_j) \mid (\forall c_i, c_j \in \mathcal{C}) \; a \in A^L(c_i) \text{ and } b \in A^L(c_j)\}$.

The scores obtained with an **Oa** reduced concept lattice is identical to those obtained with the Associative method since in an **Oa** reduced concept lattice, every domain appears only once and all proteins containing a domain appears in the extent of a concept whose intent has the domain.

## 5.2 *S. pombe* example

Interactions and non-interactions between the ten *S. pombe* proteins in the Riley dataset are depicted in Fig. 4. The presence (absence) of an edge between two nodes denotes an interaction (non-interaction). A bolded edge signifies that the two endpoint proteins form a GPPI. A GPPI is a PPI that is support by at least one GDDI.

Table 5 works out the concept-based score for the gold standard domain-pair (Pkinase, Pkinase). In the concept lattices which are not attribute-reduced, i.e.: **OA** in Fig. 2 and **oA** in Fig. SM5.1, Pkinase appears in the attribute-label set of concepts c1, c3, c2 and c7, which yields the 10 concept-pairs in Table 5. In the attribute-reduced concept lattices, i.e.: **oa** in Fig. 3 and **Oa** in Fig. SM5.2, Pkinase appears in the attribute-label set of concept c1 only and yields a single concept-pair to process.

Domain pair scoring for the four types of concept lattices: **OA**, **oA**, **Oa** and **oa**, all work in the same way. To illustrate, we take c1 × c2 from **OA**, which yields interacting protein-pairs (16, 16) and (16, 1076), and non-interacting protein-pairs (10, 16), (10, 136) and (10, 949). This makes the concept-based score for (Pkinase, Pkinase) = $\log_2(2 / (2 + 3))$. But this is not the maximum score for (Pkinase, Pkinase). For concept lattice type **OA,** the maximum score for (Pkinase, Pkinase) is $\log_2(1)$ and it is obtained through c2 × c2. We say that $CB_{OA}$(Pkinase, Pkinase) = 0.0.

$CB_{OA}$(Pkinase, Pkinase) > $CB_{Oa}$(Pkinase, Pkinase) because in **OA**, a concept lattice which is *not* attribute-reduced, Pkinase appears with other rare domains belonging to a protein, and in this case the protein (16) interacts with itself. This is an instance of a promiscuous domain (Pkinase) riding piggy-back on rare domains (the Pfam-B domains) to boost its concept-based score. In **Oa**, a concept lattice which is attribute-reduced, there is no opportunity for Pkinase to piggy-back on other domains. $CB_{Oa}$(Pkinase, Pkinase) = AM(Pkinase, Pkinase), the score obtained via the Associative method.

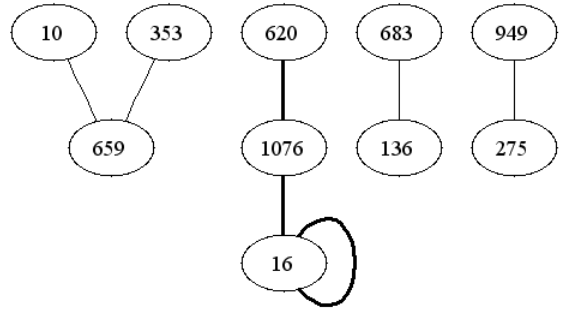

**Fig. 4** How the ten *S. pombe* proteins in the Riley dataset interact with each other. The three GPPIs are denoted with bolded edges.

**Table 5** Concept-based scoring of the GDDI(Pkinase, Pkinase) with *S. pombe* proteins and PPIs.

| Concept-pairs for GDDI(Pkinase, Pkinase) | **OA** (Fig. 2) | | **oA** (SM6) | | **Oa** (SM6) | | **Oa** (Fig. 3) | |
|---|---|---|---|---|---|---|---|---|
| | M | M+Z | M | M+Z | M | M+Z | M | M+Z |
| 1 × 1 | 2 | 15 | 0 | 3 | **2** | **15** | 0 | 3 |
| 1 × 3 | 0 | 5 | 0 | 2 | - | - | - | - |
| 1 × 2 | 2 | 5 | 1 | 2 | - | - | - | - |
| 1 × 7 | 0 | 5 | 0 | 2 | - | - | - | - |
| 3 × 3 | 0 | 1 | 0 | 1 | - | - | - | - |
| 3 × 2 | 0 | 1 | 0 | 1 | - | - | - | - |
| 3 × 7 | 0 | 1 | 0 | 1 | - | - | - | - |
| 2 × 2 | **1** | **1** | **1** | **1** | - | - | - | - |
| 2 × 7 | 0 | 1 | 0 | 1 | - | - | - | - |
| 7 × 7 | 0 | 1 | 0 | 1 | - | - | - | - |
| Concept-based score CB = max[$\log_2$(M/(M+Z))] | $\log_2(1/1)$ | | $\log_2(1/1)$ | | $\log_2(2/15)$ | | scoreless | |
| Number of score improvements (piggy-backs) PG | 2 | | 1 | | 0 | | 0 | |

The protein-pair interactions producing the higher scores for promiscuous domain-pairs need not be self-interacting; although GDDIs have a higher propensity to be self-interacting than non-GDDIs (section 3). In the *B. subtilis* example (SM6 of the Appendix), $CB_{OA}$(STAS, HATPase_c) > $CB_{Oa}$(STAS, HATPase_c) because in the **OA** concept lattice, HATPase_c appears with a rare domain (Pfam-B_8931) in c8. In the **Oa** concept lattice on the other hand, HATPase_c appears only once by definition. As such, there is no opportunity to improve the ranking for (STAS, HATPase_c). However, it is not sufficient for HATPase_c to appear with a rare domain in the **OA** concept lattice, it does this also in c5. CB$_{OA}$(STAS, HATPase_c) has the score it does because two thirds of the protein pairs in $O^L(c7) \times O^L(c8)$ are interacting. In the **OA** case, HATPase_c piggy-backs on Pfam-B_8931 and obtains a higher score for itself: CB$_{OA}$(STAS, HATPase_c) = CB$_{OA}$(STAS, Pfam-B_8931).

The two examples given before illustrate three important points:

(i) Piggy-backing on rare domains help GDDIs, which largely comprise promiscuous domains, boost their CB scores.

(ii) A necessary condition for piggy-backing is a concept lattice which is not attribute-reduced.

(iii) Another necessary condition, which is satisfied by the Riley dataset, is most proteins have mixed architecture (section 3.3). This is so that it is possible for a promiscuous domain and a rare domain to reside in the same intent.

A desirable consequence of piggy-backing is highly ranked DDIs can be enriched with GDDIs. This is one of the two tests we conduct with the entire Riley dataset in section 6. A companion test, which is also performed in section 6, is whether *the* highest ranking DDI for a GPPI is a GDDI. This method was first performed by Nye *et al*. [15] to predict domain-domain contacts for interacting protein pairs, and was used in [6] to evaluate their DDI prediction method. For brevity, we refer this method as the *Nye test*.

We use strict comparison for the Nye test: if for a GPPI, there is a non-GDDI with the same highest rank as a GDDI, the test fails for the GPPI. A GPPI with more DDIs and a small GDDI/DDI ratio poses more of a challenge for the Nye test. The protein-pair (16, 1076) is a GPPI since it is supported by the (Pkinase, Pkinase) GDDI. Following the definition in section 2, this protein-pair generates the four DDIs listed in Table 6. The GDDI (Pkinase, Pkinase) is the highest ranking DDI only when the **OA** concept lattice is used. This example illustrates the tie-breaking role of PG between some domain-pairs with identical CB scores.

**Table 6** Examining the DDIs of GPPI(16, 1076). A GDDI is the highest ranking DDI only when the **OA** concept lattice is used. PG values help differentiate domain-pairs with identical scores. PG = 0 when an attribute-reduced concept lattice is used since attribute-labels appear only once and thus there is no chance for domains to ride piggy-back.

| DDIs for GPPI(16, 1076) | **OA** | | **oA** | | **Oa** | | **oa** | |
|---|---|---|---|---|---|---|---|---|
| | **CB** | **PG** | **CB** | **PG** | **CB** | **PG** | **CB** | **PG** |
| (Pkinase, Pfam-B_106217) | 0 | 1 | 0 | 1 | -1.32193 | 0 | -1 | 0 |
| (Pkinase, Pfam-B_33993) | 0 | 1 | 0 | 1 | -1.32193 | 0 | -1 | 0 |
| (Pkinase, Pfam-B_78326) | 0 | 1 | 0 | 1 | -1.32193 | 0 | -1 | 0 |
| (Pkinase, Pkinase) | 0 | 2 | 0 | 1 | -2.90689 | 0 | scoreless | 0 |
| GDDI (Pkinase, Pkinase) is the highest ranked DDI? | Yes | | No | | No | | No | |

There are three GPPIs for *S. pombe* in the Riley dataset are listed in Table 7. Only rankings made with the **OA** concept lattice correctly placed a GDDI as the highest ranking DDI for all three GPPIs. The other three concept lattice types correctly placed a GDDI as the highest ranking DDI for only one of the GPPIs, and this GPPI has a GDDI/DDI ratio of 1.0.

**Table 7** Only rankings made with the **OA** concept lattice passes the Nye test for all three GPPIs.

| GPPI | Number of DDIs | GDDI | **OA** | **oA** | **Oa** | **oa** |
|---|---|---|---|---|---|---|
| (1076, 16) | 4 | (Pkinase, Pkinase) | Yes | No | No | No |
| (1076, 620) | 2 | (Pkinase, Cyclin_C) (Pkinase, Cyclin_N) | Yes | Yes | Yes | Yes |
| (16, 16) | 10 | (Pkinase, Pkinase) | Yes | No | No | No |
| Number of GPPIs with a GDDI as the highest ranking DDI | | | 3 | 1 | 1 | 1 |

**6. Application of concept-based domain-pair ranking to the Riley dataset**

The concept-based scoring method (section 5.1) is applied to the Riley dataset (section 3) using all four types of concept lattices (section 4.1) and the resultant domain-pair ranking is evaluated on two fronts:

(i) GDDI recovery (section 6.1), and

(ii) the Nye test (section 6.2).

The hypothesis is that when the necessary conditions for concept-based scoring are present (section 5.2), highly ranked domain-pairs will be enriched with GDDIs. This in turn will aid GDDI recovery and increase the pass rate with the Nye test.

The concept-based scoring method is further evaluated under the following four circumstances, all of which are related to the input data:

(i) Under the default or original circumstance, the complete Riley dataset is used without modification. Since all PPIs in the given input data are used for the evaluation of domain-pairs, the probability of including a PPI, $P_e = 1.0$.

(ii) PPI data obtained via high-throughput methods are error-prone. To account for the inaccuracies in PPI data, the robustness of computational methods when dealing with PPIs is commonly tested by using $P_e < 1.0$. Ref. [6] for example, reported the results for their method at $P_e = 0.5$. We do the same and evaluate the concept-based scoring method at $P_e = 0.5$, that is each PPI from the set of PPIs in the Riley dataset is included with 50% probability. Naturally, the final input data will have a reduced number of GDDIs and GPPIs.

(iii) Another way we modify the input PPI network is by permuting the nodes. We do this in such a way that the original number of PPIs per organism remains invariant and the original network structure (e.g. degree distribution, average path length, clustering and assortativity) is preserved.

(iv) Finally, to test the assertion put forth so far that a mixed domain architecture is a necessary condition for piggy-backing, we shuffled the domains of proteins such that the number of domains and their frequency of occurrence in proteins remain unchanged. We also preserve the number of domains per protein, although for this scenario we allow domain repetitions and so a protein becomes a multi-set of domains. Domain shuffling was accomplished with the following steps:

1. Place every instance of a domain in the input data into a sequence, sorted by frequency of occurrence. Domain instances with identical frequency are shuffled amongst themselves.
2. Sort the proteins by their size, i.e. number of domains they contain.
3. Starting from the largest to the smallest protein, assign domains to proteins starting from the least to the most frequently occurring domains. The reason for this is to reduce domain repetition within a protein.

The impact of shuffling on domain architecture is shown in Fig. 5 which plots the minimum and maximum domain occurrence for each protein. For example, the domain set for protein 949, D(949) = {PBD, Pfam-B_2441, Pkinase} (Table 4). The minimum and maximum occurrence values for D(949) are 1 and 5 respectively. Points from the original domain architecture (prior to domain shuffling) concentrate on the left half of the plot, while points from the shuffled (mutated) domain architecture occupy the x=y line. This reveals that mixed domain architecture in proteins is destroyed by the shuffling. The homogeneity in the mutated proteins is because protein sizes are shorter than the length of a subsequence of domains with identical occurrence. Domain shuffling changes the relation between proteins and domains. As such, a new concept lattice needs to be computed for the new relation.

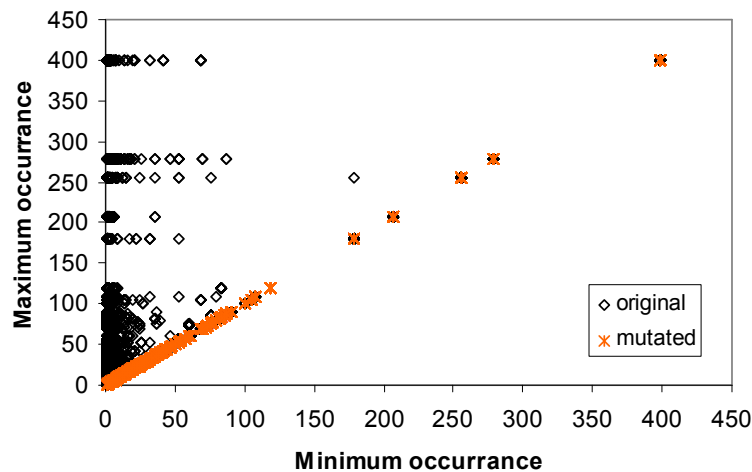

**Fig. 5** Impact of domain shuffling on domain architecture. Domain shuffling changes the original heterogeneous domain architecture to a homogeneous one in terms of domain occurrence.

## 6.1 GDDI Recovery

GDDI recovery is performed by searching domain-pairs in rank order starting from the highest to the lowest ranked domain-pairs. The search space of domain-pairs is created by generating all possible domain-pairs (DDIs) for the given set of PPIs. Domain-pairs with the same rank are shuffled amongst themselves. Each time a GDDI is found, the True Positive Rate (TPR) increases. TPR($t$) = number of GDDIs found so far at search step $t$ / total number of GDDIs to find. Each time a non-GDDI is met, the False Positive Rate (FPR) increases. FPR($t$) = number of non-GDDIs met so far at search step $t$ / total number of non-GDDIs. The total number of non-GDDIs is the number of DDIs less the number of GDDIs. Results for GDDI recovery under the four input data conditions described before are reported in Figs. 6-9. We make the following observations about the GDDI recovery results:

(i) Concept lattices that are not attribute-reduced (**OA** and **oA**) produce better results (higher TPR or Sensitivity) at lower FPR (higher Specificity).

(ii) The Associative method (**Oa**) performed the worst in all four circumstances.

(iii) The **OA** result is most robust to the changes in input data introduced here. This contradicts our expectation with regards to the impact of domain architecture and the importance of piggy-backing for our concept-based scoring method (Fig. 9). Unexpectedly also, the Associative method (**Oa**) results suffer the most from domain shuffling. However, results for the Nye test (Table 8) present an opposing view, one more in-line with our expectation. Result for the Nye test with **Oa** actually show an improvement when domains are shuffled. An improvement in GDDI recovery with **Oa** can also be seen if the **Oa** domain-pair ranks are reversed (Oa_r plot in Fig. 9). When the flipped ROC for **Oa** is taken into account, we conclude that except for domain shuffling, GDDI recovery is most robust with concept-based scoring using the **OA** concept lattice.

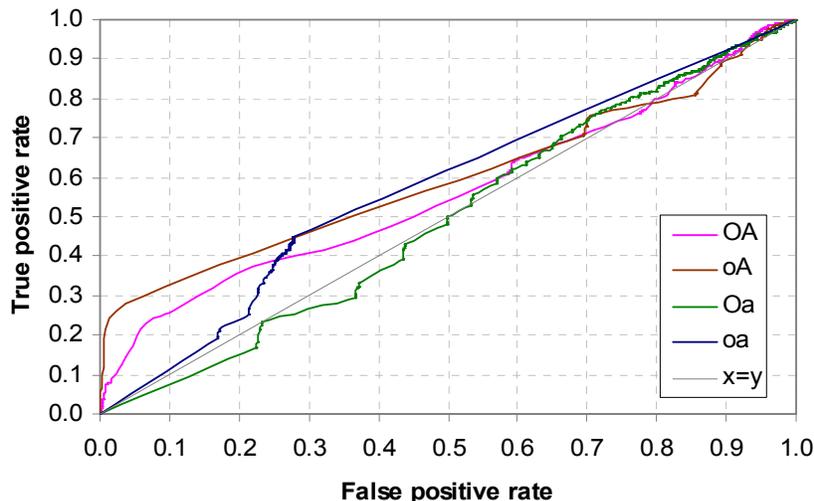

**Fig. 6** GDDI recovery under the default circumstance where Pe = 1.0

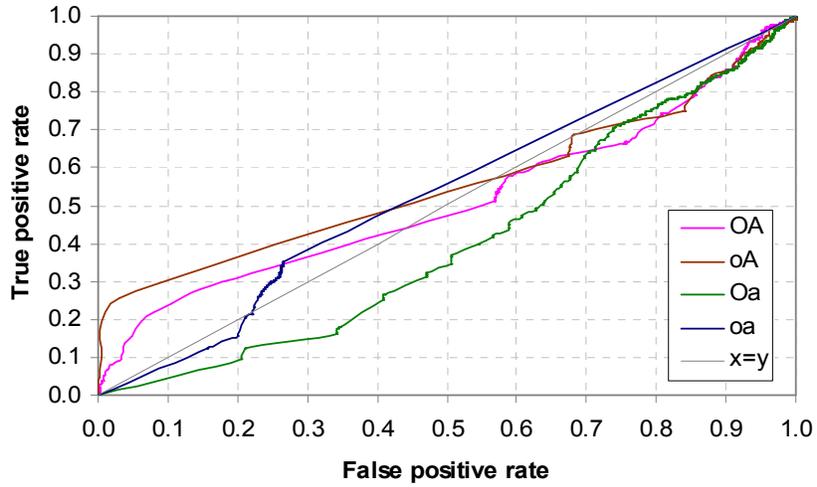
**Fig. 7** GDDI recovery with Pe = 0.5

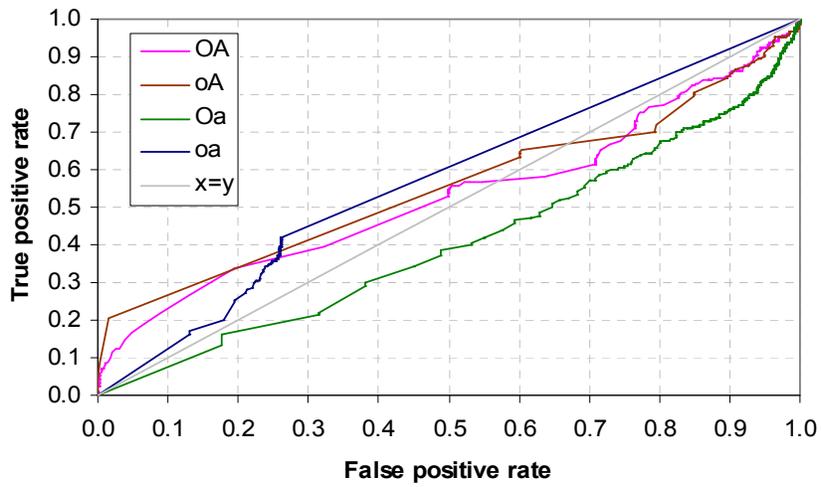
**Fig. 8** GDDI recovery with protein shuffled, Pe = 1.0

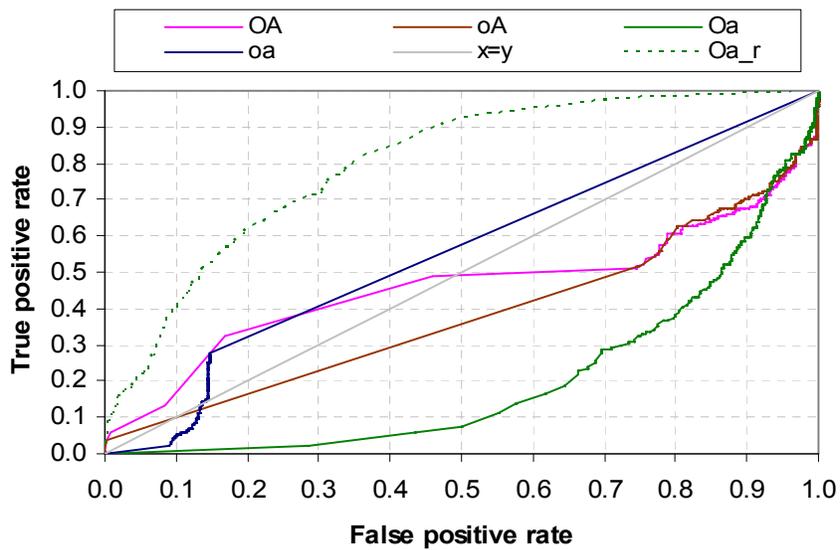
**Fig. 9** GDDI recovery with domain architecture shuffled, Pe = 1.0

## 6.2 Nye test

The Nye test [15] (section 5.2) asks of each GPPI whether *the* highest scoring DDI from all domain-pairs generated by a GPPI is a GDDI. If the highest scoring domain-pair for a GPPI is a GDDI (and there are no other non-GPPIs with the same highest score), the GPPI passes the Nye test.

Table 8 compares the percentage of GPPIs that pass the Nye test under the four test conditions. The concept lattices that are not attribute-reduced (**OA** and **oA**) produce significantly better results (higher pass or success percentage) than the attribute-reduced concept lattices (**Oa** and **oa**). This trend is consistent across all test conditions except when domains are shuffled. The Associative method (**Oa**) scored the highest success rate when the domains are shuffled. While the success rate of the other three concept lattice types deteriorated when domains were shuffled, shuffling domains actually improved the success rate for the Associative method (**Oa**). The biggest drop in success rate was suffered by concept-based scoring using the **OA** concept lattice. This is consistent with our assertion that mixed domain architecture is a necessary condition for the success of the concept-based scoring method to infer reliable DDIs from PPIs. Except for domain shuffling, the **OA** result is most robust to the changes in input data introduced here.

**Table 8** Percentage of GPPIs that pass the Nye test under the four test conditions. Except for the situation when domains are shuffled to break the original domain architecture, concept lattices which are not attribute-reduced (**oA** and **OA**) outperform concept lattices that are attribute-reduced. This supports the notion that in addition to a concept lattice that is not attribute reduced, a mixed domain architecture is also necessary for concept-based scoring to do well.

| Concept lattice type | Pe = 1.0 | | | Pe = 0.5 |
|---|---|---|---|---|
| | Default | Shuffled proteins | Shuffled domains | Default |
| **oa** | 222/1780 = 12.47% | 20/108 = 18.52% | 10/127 = 7.87% | 118/960 = 12.29% |
| **oA** | 1033/1780 = 58.03% | 37/108 = 34.26% | 20/127 = 15.75% | 505/912 = 55.37% |
| **Oa** | 188/1780 = 10.56% | 21/108 = 19.44% | 48/127 = 37.80% | 67/893 = 7.50% |
| **OA** | 1282/1780 = 72.02% | 65/108 = 60.19% | 22/127 = 17.32% | 654/902 = 72.51% |

The Nye test is more difficult for GPPIs that generate a larger number of DDIs. Most GPPIs generate a small number of DDIs, but there are GPPIs that generate well over 40 DDIs each (Fig. 10). Fig. 11 breaks down the Default (Pe = 1.0) results in Table 8 by difficulty. As the level of difficulty increases, the attribute-reduced concept lattices (**Oa** and **oa**) become less effective. None of the GPPIs that generate more than 50 DDIs pass the Nye test when concept-based scoring uses the **oa** concept lattice. For the **Oa** concept lattice (the Associative method), the cut-off point is even earlier, at 16 DDIs. In contrast, several GPPIs with more than 50 DDIs could still pass the Nye test when concept-based scoring uses the non-attributed-reduced concept lattices (**OA** and **oA**).

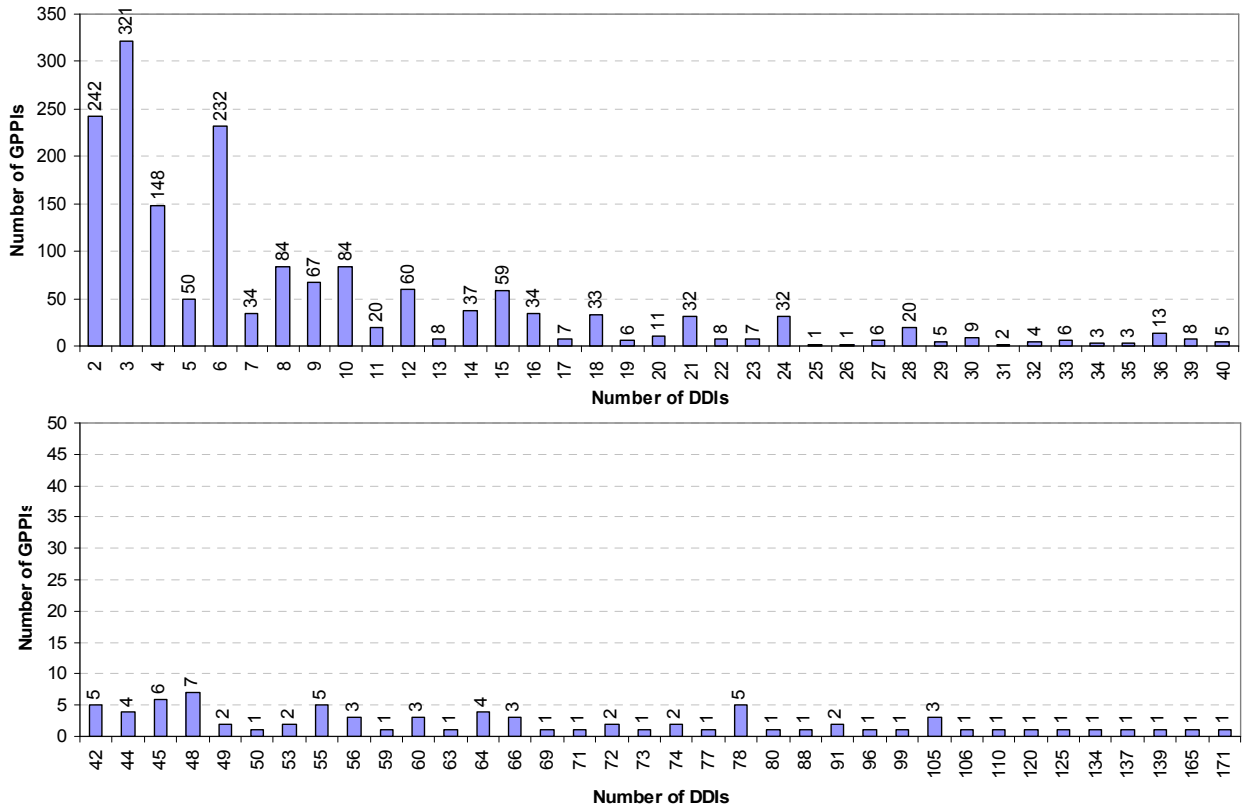

**Fig. 10** Number of GPPIs that generate *x* number of DDIs.

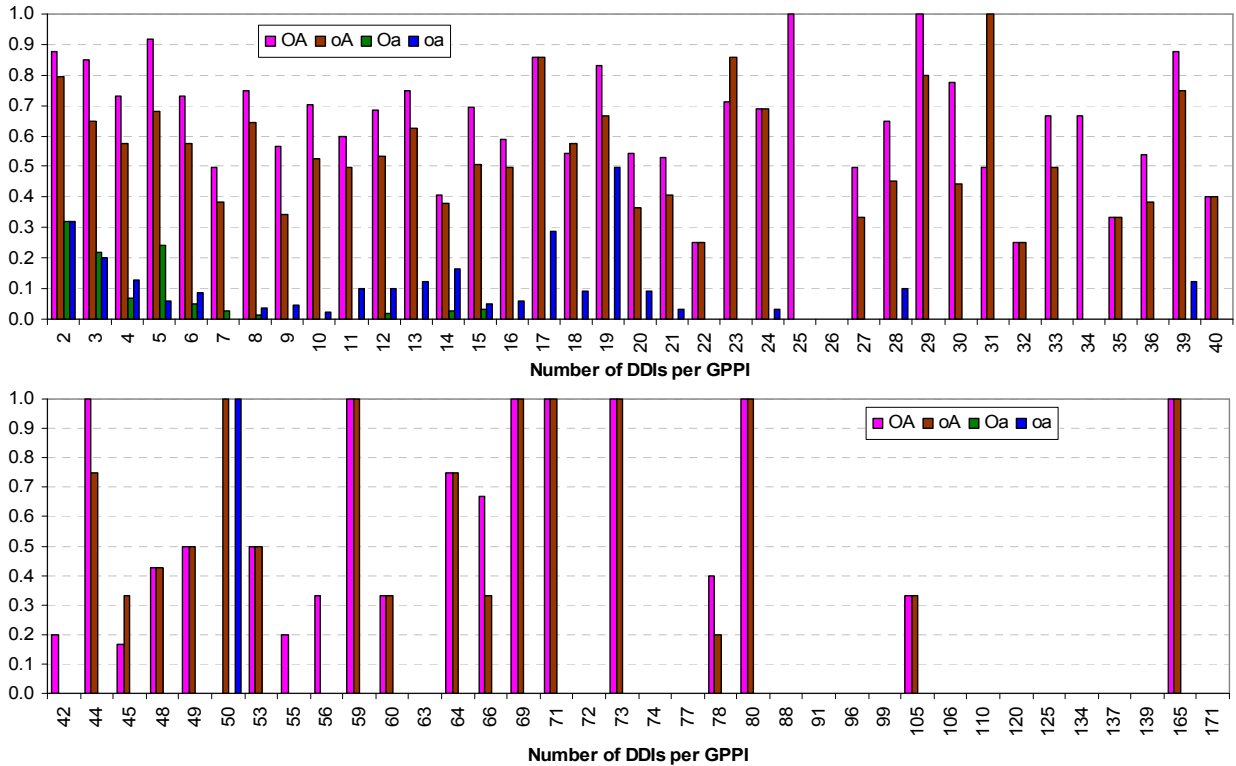

**Fig. 11** Fraction of GPPIs that pass the Nye test, i.e. where a GDDI is the highest ranking DDI. Default test condition, Pe = 1.0.

## 7. Summary and concluding remarks

A method based on Formal Concept Analysis to infer reliable domain-domain interactions from protein-protein interactions was proposed and shown to be feasible even in the presence of domain promiscuity. The proposed method outperforms the Associative method on two fronts: GDDI recovery and the Nye test. The effectiveness of the method is due to a piggy-backing mechanism. Two necessary conditions to enable piggy-backing are: (i) mixed domain architecture, and (ii) concept-based scoring and ranking based on a concept lattice that is not attribute-reduced. The problem of using highly reliable domain-pairs to predict protein-protein interactions with high accuracy remains a future challenge.

## Acknowledgement

This work was made possible by the facilities of the Shared Hierarchical Academic Research Computing Network (SHARCNET:www.sharcnet.ca) and Compute/Calcul Canada.

**Appendix (Supplementary Material)**

**SM1:** The number of common domains decreases as the number of proteins common to the training and test sets decreases.

**Table SM1**

| 11 | Park [5] | | | Analyzed with the Riley dataset | | | | | |
|----|----------|------|-----------|----------|------|-----------|---------|------|-----------|
|    | Proteins | | | Proteins | | | Domains | | |
|    | Train | Test | Intersect | Train | Test | Intersect | Train | Test | Intersect |
| CV | 2194 | 1444 | 1444 | 1893 | 1260 | 1260 | 1634 | 1206 | 1206 |
| C1 | 2194 | 1514 | 1514 | 1893 | 1313 | 1313 | 1634 | 1211 | 1211 |
| C2 | 2194 | 642  | 399  | 1893 | 581  | 362  | 1634 | 639  | 506  |
| C3 | 2194 | 424  | 0    | 1893 | 381  | 0    | 1634 | 448  | 216  |

The analysis in Table SM1 uses the training and test sets numbered 11 from [5]. According to the design in [5], C1, C2 and C3 are test sets for the same training set. CV has its own training set. We consider only proteins with domains in the Riley dataset. E.g.: Only 581 of the 642 proteins in yeast_random_test_C2_11.txt have domains in the Riley dataset. When pooled, the domains of these 581 proteins amount to 639 unique domains, 506 of which are also found in the set of domains pooled from the 1,893 training set proteins.

**SM2: Statistics for section 3.1.**

gd_freq$Freq is the frequency of occurrence for gold domains. df$freq is the frequency of occurrence for all domains.

```
> summary(gd_freq$Freq)
Min. 1st Qu.  Median    Mean 3rd Qu.    Max.
1.00    2.00    4.00   11.46   10.00  399.00
> sd(gd_freq$Freq)
[1] 26.33095

> summary(df$Freq)
Min. 1st Qu.  Median    Mean 3rd Qu.    Max.
1.000   1.000   1.000   2.488   2.000 399.000
> sd(df$Freq)
[1] 7.323162

> wilcox.test(gd_freq$Freq, df$Freq, alt="greater")
        Wilcoxon rank sum test with continuity correction
data:  gd_freq$Freq and df$Freq
W = 6291955, p-value < 2.2e-16
alternative hypothesis: true location shift is greater than 0
```

**SM3: Statistics for section 3.2**

gddi$pairs is a vector representing the number of PPIs generated per GDDI for all GDDIs. ddi$pairs is a vector representing the number of PPIs generated per DDIs for all DDIs. gddi$non_pairs is a vector representing the number of non-PPIs generated per GDDI for all GDDIs. ddi$non_pairs is a vector representing the number of non-PPIs generated per DDIs for all DDIs.

```
> summary(gddi$pairs)
   Min. 1st Qu.  Median    Mean 3rd Qu.    Max.
   1.00    1.00    2.00    4.17    4.00  120.00
> sd(gddi$pairs)
[1] 9.182882

> summary(ddi$pairs)
   Min. 1st Qu.  Median    Mean 3rd Qu.    Max.
  1.000   1.000   1.000   1.233   1.000 145.000
> sd(ddi$pairs)
[1] 1.125483

> t.test(gddi$pairs, ddi$pairs, alt="greater")
        Welch Two Sample t-test
data:  gddi$pairs and ddi$pairs
t = 8.9483, df = 782.104, p-value < 2.2e-16
alternative hypothesis: true difference in means is greater than 0
95 percent confidence interval:
 2.396194      Inf
sample estimates:
mean of x mean of y
 4.169860  1.233216

> wilcox.test(gddi$pairs, ddi$pairs, alt="greater")
        Wilcoxon rank sum test with continuity correction
data:  gddi$pairs and ddi$pairs
W = 102377372, p-value < 2.2e-16
alternative hypothesis: true location shift is greater than 0
```

**Fig. SM3.1:** Tests to confirm that GDDIs generate significantly more true positive PPIs than non-GDDIs.

```
> summary(gddi$non_pairs)
   Min. 1st Qu.  Median    Mean 3rd Qu.    Max.
    0.0     2.0     9.0   190.1    38.0 17040.0
> sd(gddi$non_pairs)
[1] 1031.264

> summary(ddi$non_pairs)
   Min. 1st Qu.  Median    Mean 3rd Qu.    Max.
   0.00    1.00    4.00   51.61   23.00 24700.00
> sd(ddi$non_pairs)
[1] 278.0015

> t.test(gddi$non_pairs, ddi$non_pairs, alt="greater")
        Welch Two Sample t-test
data:  gddi$non_pairs and ddi$non_pairs
t = 3.7561, df = 782.504, p-value = 9.269e-05
alternative hypothesis: true difference in means is greater than 0
95 percent confidence interval:
 77.74937      Inf
sample estimates:
mean of x mean of y
190.05619  51.60514

> wilcox.test(gddi$non_pairs, ddi$non_pairs, alt="greater")
        Wilcoxon rank sum test with continuity correction
data:  gddi$non_pairs and ddi$non_pairs
W = 79464083, p-value = 1.229e-13
alternative hypothesis: true location shift is greater than 0
```

**Fig. SM3.2:** Tests to confirm that GDDIs generate significantly more false positive PPIs than non-GDDIs.

**SM4: Concept lattice example to support section 4.2.**

Table SM4.1: Cross-table for the toy example in Fig. 1 [8].

|  |  | Attributes ||||||| Number of attributes per object |
|---|---|---|---|---|---|---|---|---|---|
|  |  | Y | B | A | O | G | R | V |  |
| Objects | 0 | × | × |  |  |  |  |  | 2 |
|  | 1 |  | × | × | × | × | × |  | 5 |
|  | 2 |  |  |  |  |  | × |  | 1 |
|  | 3 |  |  |  | × |  | × |  | 2 |
|  | 4 | × | × |  |  | × | × | × | 5 |
|  | 5 |  |  |  | × |  | × |  | 2 |
|  | 6 |  | × |  |  |  |  |  | 1 |
|  | 7 |  | × |  |  |  |  |  | 1 |
|  | 8 |  |  |  |  | × | × |  | 2 |
|  | 9 |  |  |  |  |  | × |  | 1 |
|  | 10 |  | × | × | × | × |  | × | 5 |
|  | 11 |  |  |  |  |  | × |  | 1 |
| Attribute Frequency |  | 2 | 6 | 2 | 4 | 4 | 8 | 2 |  |

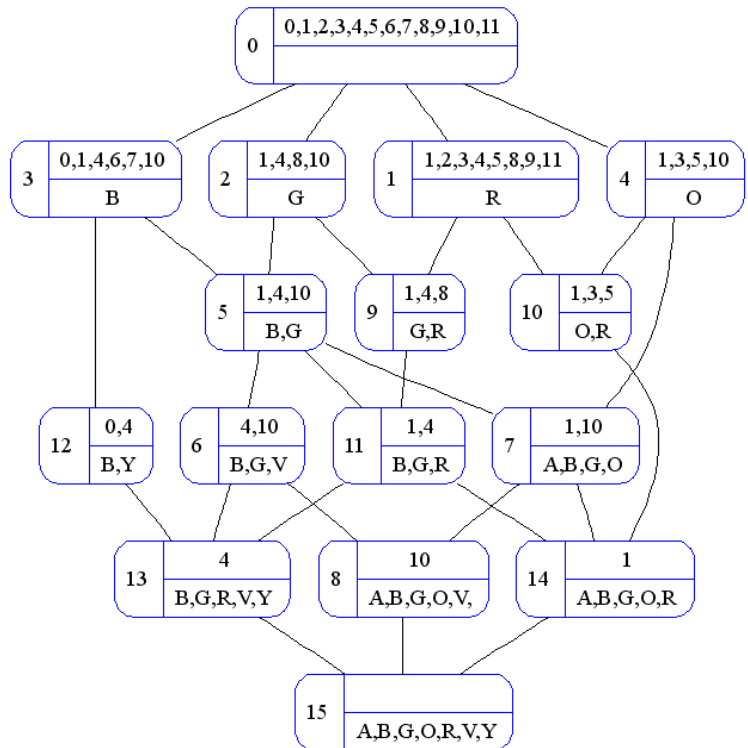

**Fig. SM4.1:** The **OA** concept lattice for the relation in Table SM4.1

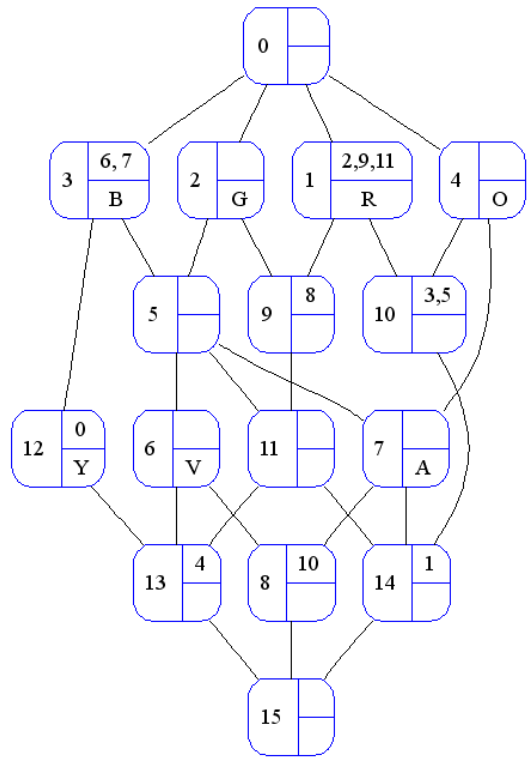

**Fig. SM4.2:** The **oa** concept lattice for the relation in Table SM4.1

Attributes with frequency > 4 (a.k.a. the promiscuous attributes: B, G, R and O) appear in concepts one step away from the top concept. The remaining attributes Y, V and A all appear only twice in the set of objects, and they appear at least two steps away from the top concept. Promiscuous attributes appear towards the top and rare attributes appear towards the bottom of an attribute-reduced concept lattice.

Objects comprising only one frequently occurring attributes (6, 7, 2, 9 and 11) appear in concepts one step away from the top concept. Objects with two or more promiscuous attributes only (0, 8, 3, 5) appear in concepts at least two steps away from the top concept. The objects comprising promiscuous and rare attributes (mixed attribute architecture: 4, 10, 1) appear one step away from the bottom concept. Objects comprising only promiscuous attributes appear towards the top of an object-reduced concept lattice. Mixed attribute objects appear towards the bottom of an object-reduced concept lattice.

**SM5: Partially reduced concept lattices for *S. pombe***

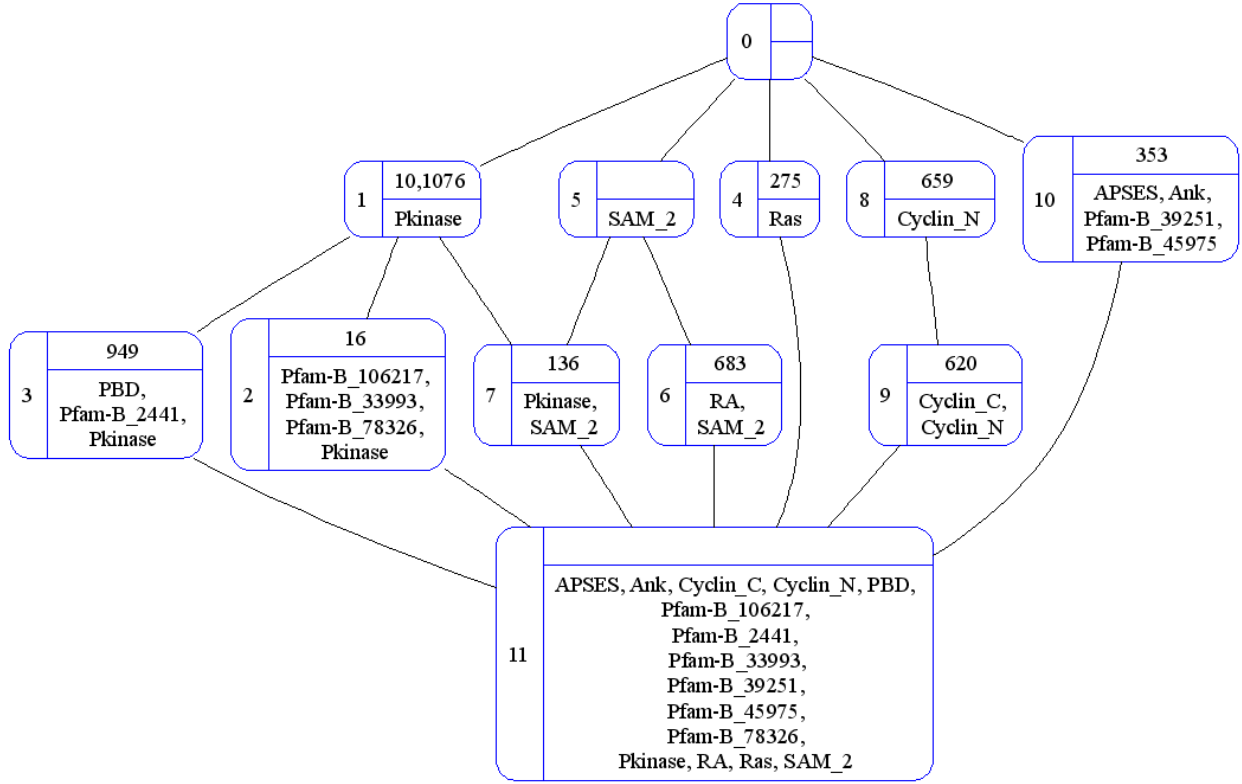

**Fig. SM5.1** The **oA** concept lattice for the *S. pombe* relation in Table 4.

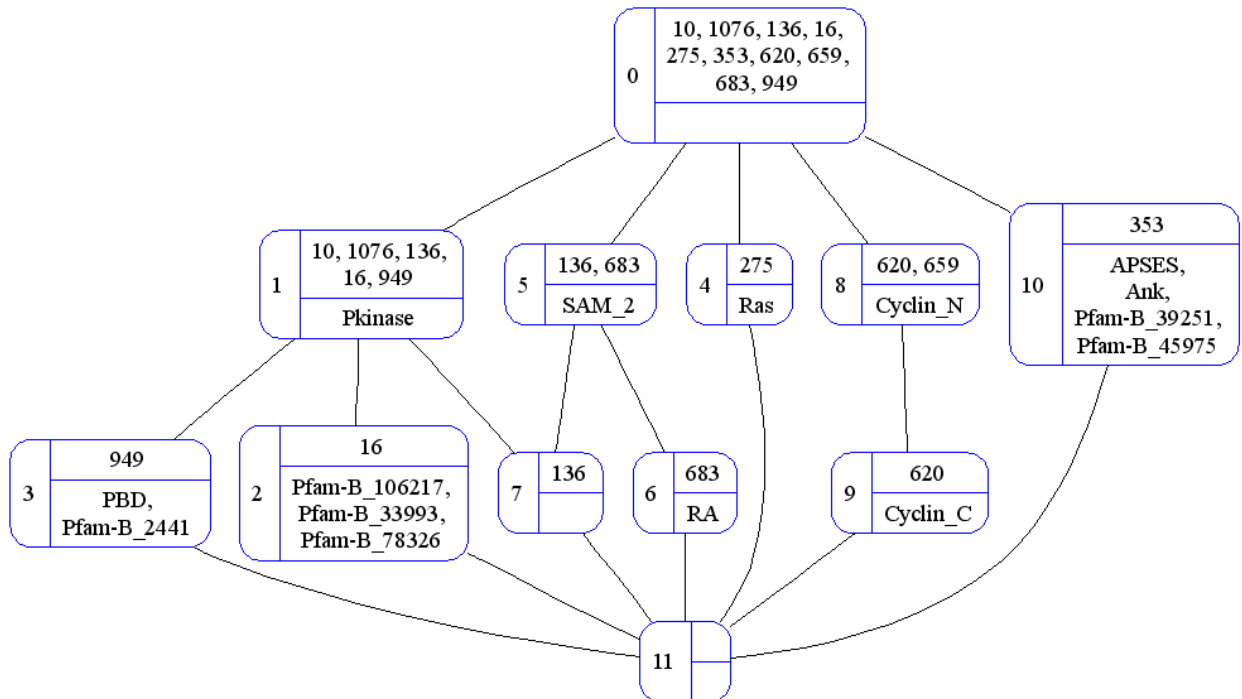

**Fig. SM5.2** The **Oa** concept lattice for the *S. pombe* relation in Table 4.

**SM6: A demonstration of concept-based scoring with *Bacillus subtilis***

**Table SM6.1**: A cross-table representing the relation between proteins and domains associated with *B. subtilis* in the Riley dataset. E.g.: the domain set for protein 402, D(402) = {HATPase_c, Pfam-B_8931}; and the protein set for domain STAS, P(STAS) = {375, 382, 92}.

|  |  | Objects = Proteins (uid) | | | | | | | | | Domain |
|---|---|---|---|---|---|---|---|---|---|---|---|
|  |  | 319 | 375 | 382 | 402 | 409 | 410 | 6102 | 6103 | 92 | Freq. |
| Attributes = Domains | HATPase_c | × |  |  | × |  |  |  |  |  | 2 |
|  | HTH_3 |  |  |  |  |  |  | × |  |  | 1 |
|  | Mn_catalase |  |  |  |  | × |  |  |  |  | 1 |
|  | Pfam-B_21839 |  |  |  |  |  | × |  |  |  | 1 |
|  | Pfam-B_3091 | × |  |  |  |  |  |  |  |  | 1 |
|  | Pfam-B_32775 |  |  |  |  |  |  | × |  |  | 1 |
|  | Pfam-B_8931 |  |  |  | × |  |  |  |  |  | 1 |
|  | Pfam-B_92151 |  |  |  |  |  |  |  | × |  | 1 |
|  | STAS |  | × | × |  |  |  |  |  | × | 3 |
|  | SpoIIE |  |  |  |  |  | × |  |  |  | 1 |
| Domains per protein | | 2 | 1 | 1 | 2 | 1 | 2 | 2 | 1 | 1 | |

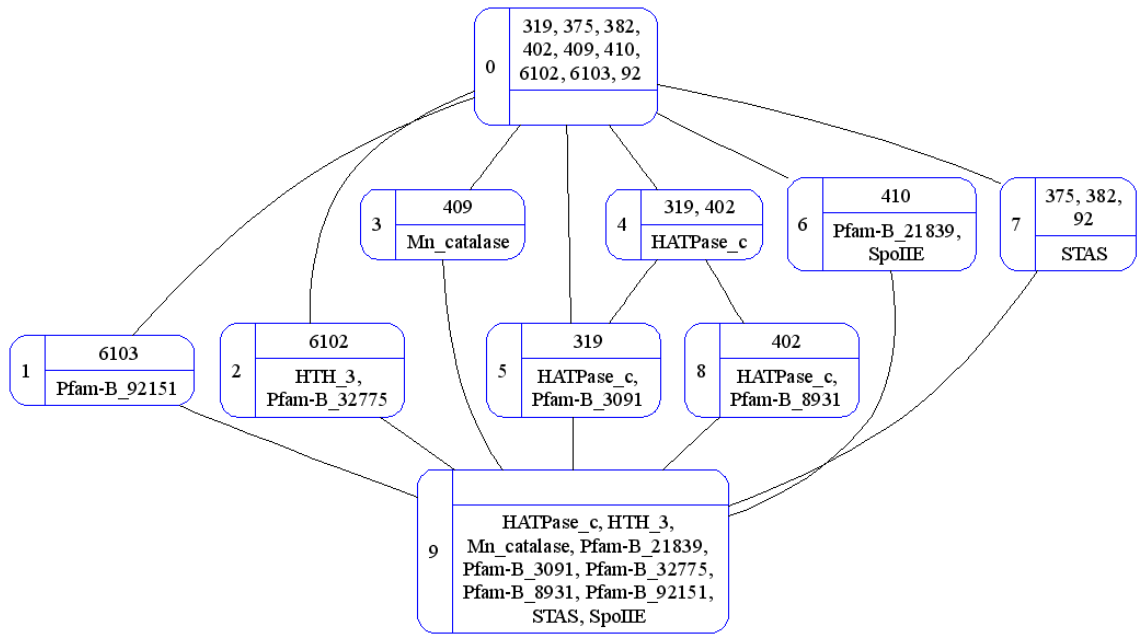

**Fig. SM6.1** The **OA** concept lattice for the *B.subtilis* relation in Table SM6.1.

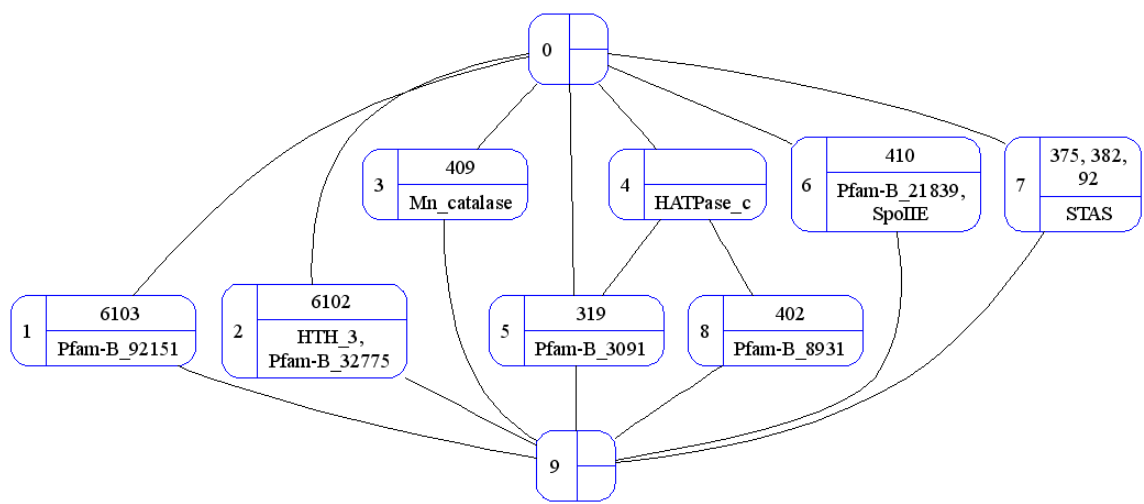

**Fig. SM6.2** The **oa** concept lattice for the *B.subtilis* relation in Table SM6.1.

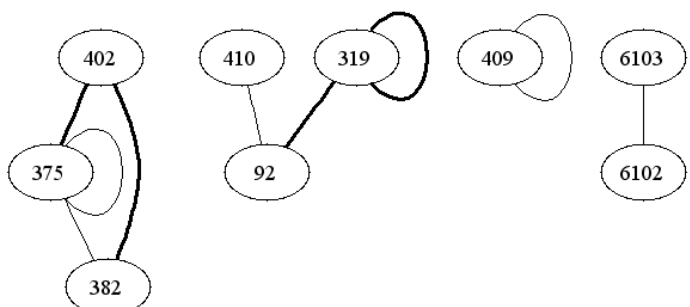

**Fig. SM6.3** Interactions between the nine *B. subtilis* proteins in the Riley dataset. The four GPPIs are denoted with bolded edges.

**Table SM6.2** Concept-based scoring of the GDDI(HATPase_c, STAS) with *B. subtilis* proteins and PPIs.

| Concept-pairs for GDDI (HATPase_c, STAS) | OA | | oA | | Oa | | oa | |
|---|---|---|---|---|---|---|---|---|
| | M | M+Z | M | M+Z | M | M+Z | M | M+Z |
| $7 \times 4$ | 3 | 6 | - | - | **3** | **6** | 0 | 0 |
| $7 \times 5$ | 1 | 3 | 1 | 3 | - | - | - | - |
| $7 \times 8$ | **2** | **3** | **2** | **3** | - | - | - | - |
| Concept-based score CB = max[$\log_2(M/(M+Z))$] | $\log_2 (2/3)$ | | $\log_2 (2/3)$ | | $\log_2 (3/6)$ | | scoreless | |
| Number of score improvements (piggy-backs) PG | 2 | | 1 | | 0 | | 0 | |

**Table SM6.3** Only rankings made with the **OA** concept lattice passes the Nye test for all four GPPIs.

| GPPI | Number of DDIs | GDDI | **OA** | oA | Oa | oa |
|---|---|---|---|---|---|---|
| (319, 319) | 3 | (HATPase_c, HATPase_c) | Yes | No | No | No |
| (319, 92) | 2 | (STAS, HATPase_c) | Yes | Yes | Yes | No |
| (375, 402) | 2 | (STAS, HATPase_c) | Yes | Yes | No | No |
| (382, 402) | 2 | (STAS, HATPase_c) | Yes | Yes | No | No |
| Number of GPPIs with a GDDI as the highest ranking DDI | | | 4 | 3 | 1 | 0 |

28